\documentclass[sigconf,authorversion=true, nonacm=true]{acmart}
\usepackage{multirow}
\usepackage{enumitem}
\usepackage{subcaption}
\usepackage{bbm}
\usepackage{graphicx}
\usepackage{mathtools}

\usepackage[linesnumbered,ruled,vlined]{algorithm2e}
\AtBeginDocument{%
  \providecommand\BibTeX{{%
    \normalfont B\kern-0.5em{\scshape i\kern-0.25em b}\kern-0.8em\TeX}}}

\setcopyright{acmcopyright}
\copyrightyear{2021}
\acmYear{2021}
\acmDOI{nn.nnnn/nnnnnnn.nnnnnnn}

\acmConference[CIKM '21]{CIKM '21: The 30th ACM International Conference on Information and Knowledge Management}{November 1-5, 2021}{Online}
\acmBooktitle{CIKM '21: The 30th ACM International Conference on Information and Knowledge Management, November 1-5, 2021, Online}
\acmPrice{15.00}
\acmISBN{XXX-X-XXXX-XXXX-X/YY/MM}

\newcommand{\modelname}{\textsf{CoSeRec}\xspace}

\setlist[itemize]{leftmargin=*}

\begin{document}

\title{Contrastive Self-supervised Sequential Recommendation with Robust Augmentation}

\author{Zhiwei Liu}
\authornote{Both authors contributed equally to this research.}
\authornote{Work done during the internship at Salesforce.}
\affiliation{%
  \institution{University of Illinois at Chicago}
}
\email{jim96liu@gmail.com }

\author{Yongjun Chen}
\authornotemark[1]
\affiliation{%
  \institution{Salesforce Research}
}
\email{yongjun.chen@salesforce.com}

\author{Jia Li}
\affiliation{%
  \institution{Salesforce Research}}
\email{jia.li@salesforce.com}

\author{Philip S. Yu}
\affiliation{%
  \institution{University of Illinois at Chicago}}
\email{psyu@uic.edu}

\author{Julian McAuley}
\affiliation{%
  \institution{UC San Diego}}
\email{jmcauley@ucsd.edu}

\author{Caiming Xiong}
\affiliation{%
  \institution{Salesforce Research}}
\email{cxiong@salesforce.com}



\begin{abstract}
\emph{Sequential Recommendation}
describes a set of techniques to model dynamic user behavior in order to predict future interactions in sequential user data.
At their core, such approaches model transition probabilities between items in a sequence, whether through Markov chains, recurrent networks, or more recently, Transformers.
However both old and new issues remain, including 
data-sparsity and noisy data;
such issues can impair performance, especially in complex, parameter-hungry models.
In this paper, we investigate the 
application
of contrastive Self-Supervised Learning~(SSL) to sequential recommendation,
as a way to
alleviate some of these issues. Contrastive SSL constructs augmentations from unlabelled instances, where agreements among positive pairs are maximized. 
It is challenging to devise a contrastive SSL framework for sequential recommendation, due to its discrete
nature,
correlations among items, and skewness of length distributions. To this end, we propose a novel framework, Contrastive Self-supervised Learning for Sequential Recommendation~(\modelname). We introduce two informative augmentation operators leveraging item correlations to create high quality views for contrastive learning. 
Experimental results on three real-world datasets demonstrate the effectiveness 
of the proposed method on improving model performance, and the robustness against
sparse and noisy data. Our implementation is available: \textsf{https://github.com/YChen1993/CoSeRec}

\end{abstract}



\keywords{Contrastive Learning, Robustness, Sequential Recommendation}



\maketitle

\section{Introduction}
Recommender systems aim at predicting 
potential interests of users toward unseen items~\cite{bpr2009rendle,liu2020basconv,wang19neural} by leveraging historical interaction records.
In order to model
dynamic 
user
behavior (and especially \emph{short-term} dynamics),
Sequential Recommendation~(SR)~\cite{rendle2010factorizing,hidasi2015session,kang2018self,sun2019bert4rec,ssept20wu} 
arguably represents the current state-of-the-art.
SR 
models interactions between users and items as temporally-ordered sequences, with the goal of predicting the next interaction conditioned on recent observations.

The core idea to model 
SR problems is to capture relationships among items in sequences%
~\cite{rendle2010factorizing,wu2017recurrent,kang2018self,hidasi2015session}. 
Pioneering works~\cite{rendle2010factorizing,he2016fusing} employ Markov chains to learn pair-wise transition 
relationships.
Later, RNN-based~\cite{hidasi2015session,wu2017recurrent,yu2016dynamic} models were proposed to infer sequence-wise correlations. 
More recently, the powerful ability of Transformers~\cite{vaswani2017attention,devlin2018bert} in encoding sequences 
has been adopted for applications in
SR~\cite{kang2018self,sun2019bert4rec,li2020time,ye2020time}. Transformer-based SR models~\cite{kang2018self,sun2019bert4rec,ssept20wu} encode sequences by learning item importances via self-attention mechanisms.

Despite the effectiveness of existing 
approaches,
several 
issues
remain less explored: 
(1) 
\textit{Data-sparsity}.
Transformers 
are motivated by
NLP tasks~\cite{vaswani2017attention}, where large
corpora
are available to train complex models~\cite{devlin2018bert}. However, SR tasks usually involve rather sparse datasets~\cite{liu2021augmenting,mcauley2015image}, which undermines the capability of a Transformer to model item correlations in sequences, \textit{e.g.} performing poorly on short sequences~\cite{liu2021augmenting}. (2) \textit{Noisy interactions}. 
Occasionally, items in sequences may not reflect 
true item correlations, \textit{e.g.} a user consumes an item because of promotion ads, 
ultimately resulting in
negative feedback~\cite{wang2021denoising}. Inference from those noisy sequences spoils the performance of a model in revealing item transition correlations, thus producing less satisfying recommendation.


Inspired by recent developments of Self-Supervised Learning (SSL)~\cite{misra2020self,lan2019albert,devlin2018bert,fang2020cert,jiao2020sub,wu2020self}, this paper studies the possibility of contrastive self-supervised learning~\cite{chen2020simple,fang2020cert,wu2020self} for alleviating the aforementioned issues in SR. Intuitively, 
SSL
constructs \textit{augmentations} from unlabelled data,
and enhances the discrimination ability of encoders by employing
a contrastive loss~\cite{zhou2020contrastive,chen2020simple,he2020momentum,you2020graph}. The contrastive loss maximizes
agreements between positive pairs, which are different augmentations (\textit{i.e.} views) from one instance.  
It is worth noting that this learning scheme is different from widely used contrastive losses in recommender systems, \textit{e.g.} BPR~\cite{bpr2009rendle} and NCE~\cite{gutmann2010noise} losses, because contrastive SSL involves no prediction targets between two different entities but only the optimization of the consistency between two views of one instance. 
Nevertheless, it is challenging to devise a contrastive SSL framework for SR due to the following reasons:
\begin{itemize}
    \item \textbf{Discreteness:} Items in sequences are represented by discrete IDs, which 
    pose a challenge for
    popular augmentation methods designed 
    for
    continuous feature spaces~\cite{chen2020simple,he2020momentum}. 
    Augmentation techniques for sequence
    data
    are
    still under-explored and require investigation. 
    
    \item \textbf{Item Correlations:} Items in sequences are correlated with each other. However, existing techniques~\cite{chen2020simple,fang2020cert,wu2020self,xie2020contrastive} augment sequences 
    based on random item perturbations. Those methods tend to 
    destroy
    the original item relationships in sequences when constructing augmentations, thus impairing the confidence of positive pairs. 
    
    %
    
    \item \textbf{Length Skewness:} According to~\cite{liu2021augmenting}, the length distribution of sequences is skewed and suffers from the long-tail patterns. However, current contrastive SSL~\cite{chen2020simple,xie2020contrastive,wu2020self} frameworks employ identical augmentation methods 
    across
    instances. 
    This
    leads to less confident contrastive pairs in SR, \textit{e.g.}, short sequences are more sensitive to 
    random item perturbations.

\end{itemize}

To this end, we propose a new framework, 
\textbf{Co}ntrastive \textbf{Se}lf-Supervised learning for \textbf{Se}quential \textbf{Rec}ommendation~(\modelname). It contains three key components:
(1) robust data augmentation that characterizes item correlations and sequence length skewness;
(2) a contrastive SSL objective to maximize the agreement of positive views of sequences; 
and (3) a multi-task training strategy to jointly optimize the SR objective 
and the contrastive SSL objective effectively. 
To be more specific, we first propose two \textit{informative augmentation} methods, 
\emph{substitute} and \emph{insert}, which advances current random augmentations~\cite{xie2020contrastive,zhou2020s3}, such as \emph{crop}, \emph{reorder}, and \emph{mask}.  
Random augmentations may yield less confident positive pairs, as those random perturbations of items break item relationships in sequences. 
Random augmentations can also hinder the quality of learned representations, because they (\textit{e.g.} crop) often lead to fewer items  in sequences, which exaggerates the cold-start issue.
Unlike random augmentations, our informative augmentations leverage item correlations when constructing positive views, thus being more robust.
Additionally, both `substitute' and `insert' operations expand user interaction records, which alleviates the cold-start issue. 
\modelname adopts a multi-task training strategy rather than a pre-training scheme~\cite{chen2020simple,wu2020clear},
because 
next-item prediction and the contrastive SSL 
share
similar targets, 
\textit{i.e.} modeling item relationships in sequences. 
Experiments on three real-world datasets verify the efficacy of \modelname. 
We summarize our contributions as follows:
\begin{itemize}
    \item We devise a novel learning framework (\modelname), which addresses data-sparsity and noisy interaction issues by unifying contrastive SSL with sequential recommendation.
    \item We propose novel and robust sequence augmentation methods that exploit item correlations and alleviate length skewness problems in sequential recommendation.
    \item We conduct extensive experiments on three benchmark datasets with detailed analyses of the proposed paradigm and the optimal combinations of augmentations. 
\end{itemize}


\section{Related Work}
\subsection{Sequential Recommendation}
\label{subsec:sr}
SR predicts future items in user 
sequences by modeling item transition correlations. Pioneering works~\cite{rendle2010factorizing,he2016fusing}  adopt Markov chains to model the pair-wise item transition correlations. FPMC~\cite{rendle2010factorizing} directly factorizes item transition matrices. Fossil~\cite{he2016fusing} extends this idea by leveraging similarities between items, which alleviates 
sparsity issues. Later, 
Recurrent Neural Network~(RNN) have been adapted to solve SR~\cite{hidasi2015session,wu2017recurrent,yu2016dynamic}, 
ostensibly modeling sequence-level correlations among transitions.
Hierarchical RNNs~\cite{quadrana2017personalizing} enhance 
RNNs 
using 
personalization
information. 
Wu, et al.~\cite{wu2017recurrent} 
apply
LSTMs to explore both 
long-term and short-term item transition correlations. 
The major drawback of 
Markov chain and RNN
models
is
that their receptive fields of the transition function are limited. Therefore, they can only incorporate shallow item correlations, which limits 
their power to encode sequences~\cite{chen2018sequential}. 


Recently, owing to the success of self-attention models~\cite{vaswani2017attention,devlin2018bert} in NLP tasks, a series of Transformer-based SR models have been proposed~\cite{kang2018self,sun2019bert4rec,ma2020disentangled,wu2020deja}. SASRec~\cite{kang2018self} applies Transformer layer to learn item importance in sequences, which 
characterize complex item transition correlations. Later, inspired by 
BERT~\cite{devlin2018bert} model, BERT4Rec~\cite{sun2019bert4rec} is proposed with a bidirectional Transformer layer.
Other works~\cite{ma2020disentangled,ssept20wu,li2020time,fan2021Modeling,fan2021continuous} also extend Transformer to incorporate complex signals in sequences, which verifies the efficacy of Transformer in solving SR. 
However, a recent work~\cite{liu2021augmenting} argues Transformer is vulnerable to 
severe user cold-start issues in SR, where existing Transformer-based models yield unsatisfying results for short sequences. Therefore, 
augmentation for short sequences is desirable. S$^3$-Rec~\cite{zhou2020s3} and CLS4Rec~\cite{xie2020contrastive} also investigate 
contrastive learning in SR. However, both 
investigate weak self-supervised signals with random augmentations in sequences, 
but do not yet address the length skewness and item correlation challenges.

\subsection{Contrastive Self-supervised Learning}

Contrastive learning 
has recently achieved
remarkable successes 
when
coupled with the Self-Supervised Learning~(SSL) framework, 
in areas ranging from
Computer Vision~(CV)~\cite{chen2020simple,he2020momentum}, Natural Language Understanding~(NLU)~\cite{fang2020cert,gao2021simcse,wu2020clear}, graph embedding~\cite{you2020graph,jiao2020sub,zhu2020graph}, as well as recommender systems~\cite{wu2020self,zhou2020s3,xie2020contrastive}. Contrastive SSL trains encoders by maximizing 
agreement between two augmented views of one instance, \textit{i.e.}, 
positive pairs. As such, it encourages self-supervised training from 
the space of
unlabelled data.
The core of 
Contrastive SSL
is to exploit strong augmentations~\cite{xiao2020should} of data instances, which are diverse with respect to the data domains and objective tasks. 

SimCLR~\cite{chen2020simple} proposes a simple contrastive learning framework for visual representations, which studies stochastic image augmentations. Following 
this,
\cite{tian2020makes,xiao2020should} argues that the optimal choice of augmentations is critically task-oriented and exists with balanced mutual information. 
As 
for
NLU tasks, the augmentation~\cite{wei2019eda} for text is of 
a
discrete nature. Deleting~\cite{wu2020clear,giorgi2020declutr}, reordering~\cite{wu2020clear} and substituting~\cite{fang2020cert,wu2020clear} words in a sentence leads to positive augmentation pairs, which also inspires this work. The augmentation in graph contrastive SSL differs from other domains in its structural characteristics. Thus, node-drop~\cite{you2020graph,zhu2020graph}, edge-drop~\cite{you2020graph,zhu2020graph}, and random walk~\cite{qiu2020gcc,velivckovic2018deep} are adopted as the augmentation operations. 

In 
recommendation
scenarios, contrastive learning is not a new technology, \textit{e.g.}, the widely used BPR loss~\cite{rendle2012bpr} and NCE loss~\cite{gutmann2010noise}, both 
adopt
pair-wise contrastive losses on positive and negative samples. 
Their
recent union with SSL 
shows promise
in improving recommender systems~\cite{wu2020self}.
S$^3$-Rec~\cite{zhou2020s3} devises contrastive SSL to maximize the mutual information over attributes and sequence augmentations, which adopts random masks of attributes and items. \cite{yao2020self} proposes two-stage augmentations by masking item embedding layers and dropping categorical features, which enables SSL for item encoders. SGL~\cite{wu2020self} advances graph-based recommender systems with SSL by employing graph structure augmentations. A 
recent work, CLS4Rec~\cite{xie2020contrastive} proposes a contrastive SSL framework for improving SR. 
Although 
close to our
paper, 
CLS4Rec
adopt 
random augmentation methods, 
ignoring
crucial item correlations and length skewness for sequence augmentations. 

\section{Preliminaries}
\begin{figure*}
    \centering
    \includegraphics[width=0.8\textwidth]{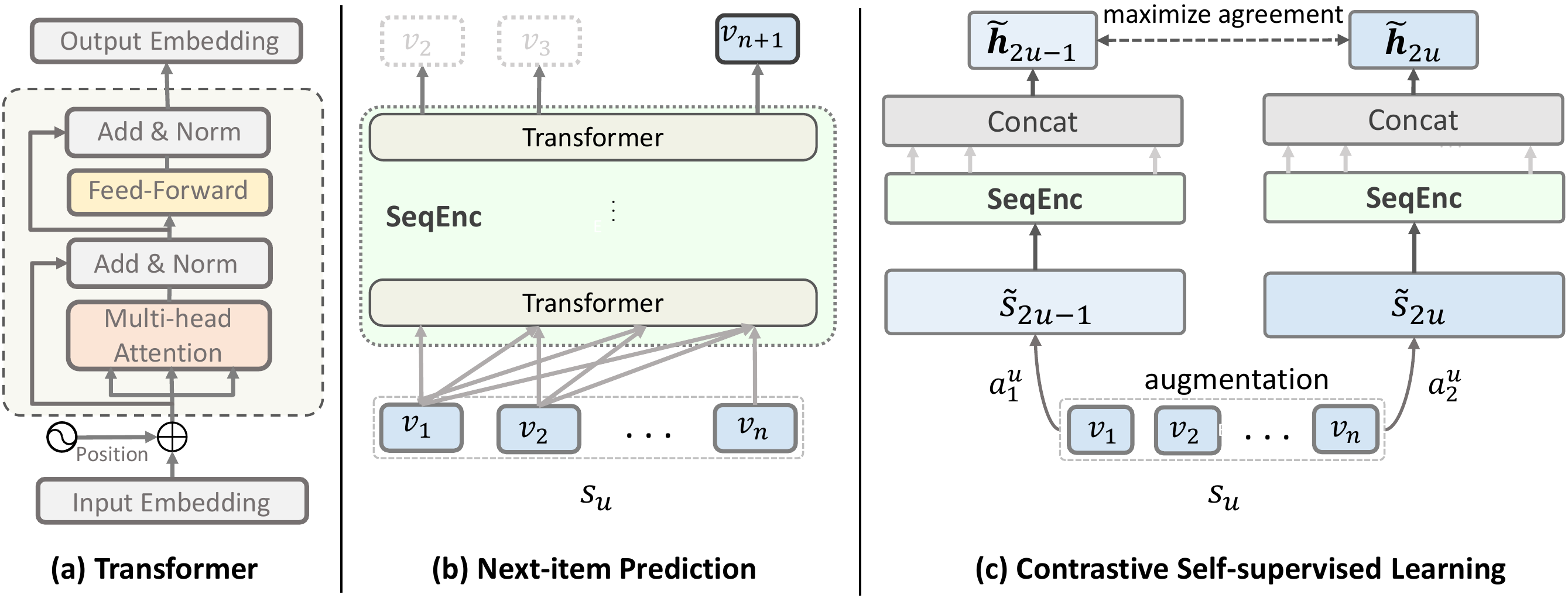}
    \caption{Overall framework. (a) illustrates the structure of Transformer. (b) presents the sequence encoding and prediction of Transformer; $v_i$ denotes an item at position $i$ of the sequence $s_u$. (c) demonstrates the contrastive SSL of a sequence. It first augments a sequence as two views with two augmentation methods. Then, it encodes the sequence by concatenating embedding outputs from Transformer. Finally, it maximizes the agreement between the two views. }
    \label{fig:framework}
\end{figure*}
\subsection{Problem Formulation}
We denote 
user and item sets as $\mathcal{U}$ and $\mathcal{V}$ respectively. Each user $u\in \mathcal{U}$ is associated with a sequence of items in chronological order
$s_{u}= [v_{1}, \dots, v_{t}, \dots, v_{|s_{u}|}]$, where $v_{t}\in \mathcal{V}$ denotes the item that $u$
has interacted with at time $t$ and $|s_{u}|$ is the total number of items. 
SR tasks 
seek to predict
the next item $v_{|s_{u}|+1}$, which is formulated as follows:
\begin{equation}
\underset{v_{i}\in \mathcal{V}}{\mathrm{arg\,max}}~P(v_{|s_{u}|+1}=v_{i}\left| s_{u}\right.),
\end{equation}
which is interpreted as calculating the probability of all candidate items and selecting the highest one for recommendation.

\subsection{Transformer for SR}\label{sec:transformer-encoder}
Transformer~\cite{vaswani2017attention} architecture 
is a powerful way to encode sequences,
leading to successful application in
SR~\cite{kang2018self,li2020time,zhou2020s3}.
It consists of two key components:
a \textit{multi-head self-attention module} and a position-wise \textit{Feed-Forward Network}~(FFN).
We illustrate the structure of a Transformer in Figure~\ref{fig:framework}(a). Multi-head self-attention models the item correlations in sequence. Position-wise FFN outputs a bag of embeddings, where the embedding at each position predicts the corresponding next item in the sequence. We present 
next-item prediction in Figure~\ref{fig:framework}(b). To keep generality, we denote a Transformer encoder as $\mathsf{SeqEnc}(\cdot)$, which 
can be an arbitrary sequence encoder. We formulate the encoding process as: 
\begin{equation}\label{eq:seq_encoder}
\mathbf{h}_{u} = \mathsf{SeqEnc}(s_{u}),
\end{equation}
where $\mathbf{h}_{u}$ denotes the sequence embedding of $s_{u}$. $\mathbf{h}_{u}$ is a bag of embeddings in Transformer. 
For
each position $t$,
$\mathbf{h}_{u}^{t}$, represents a predicted next-item~\cite{kang2018self}.
We adopt the log-likelihood loss function to optimize the encoder for next-item prediction as follows:
\begin{equation}
\mathcal{L}_{\text{rec}}(u,t) = -\log(\sigma(\mathbf{h}_{u}^{t}\cdot \mathbf{e}_{v_{t+1}}))- \sum_{v_{j}\not\in s_{u}}\log(1-\sigma (\mathbf{h}_{u}^{t} \cdot \mathbf{e}_{v_{j}})),
\end{equation}
where $\mathcal{L}_{\text{rec}}(u,t)$ denotes the loss score for the prediction at position $t$ in sequence $s_{u}$, $\sigma$ is the non-linear activation function, $\mathbf{e}_{v_{t+1}}$ denotes the embedding for item $v_{t+1}$, and $v_{j}$ is the sampled negative item for $s_u$. 
The embeddings of items are retrieved from the embedding table in $\mathsf{SeqEnc}$, which is jointly optimized with Transformer. 

\section{Methodology}

We first introduce data augmentations adopted in this paper, including both random and informative augmentations. Then we describe contrastive SSL with those augmentations. Finally, we present the overall training algorithm. 


\subsection{Robust Sequence Augmentation}
\begin{figure}
\begin{subfigure}[b]{0.22\textwidth}
    \includegraphics[width=0.9\textwidth]{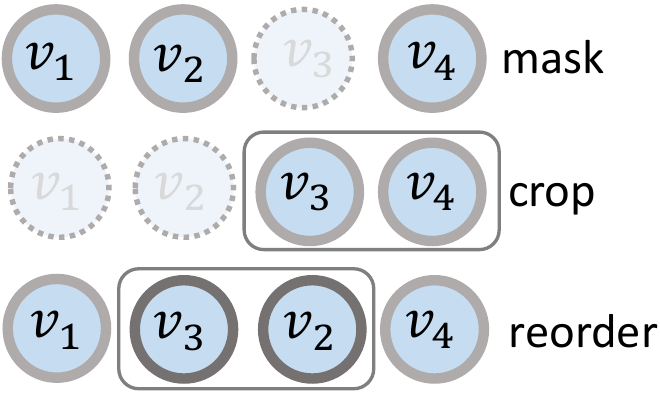}
    \caption{Random}
    \label{fig:random_aug}
\end{subfigure}
\begin{subfigure}[b]{0.25\textwidth}
    \includegraphics[width=0.9\textwidth]{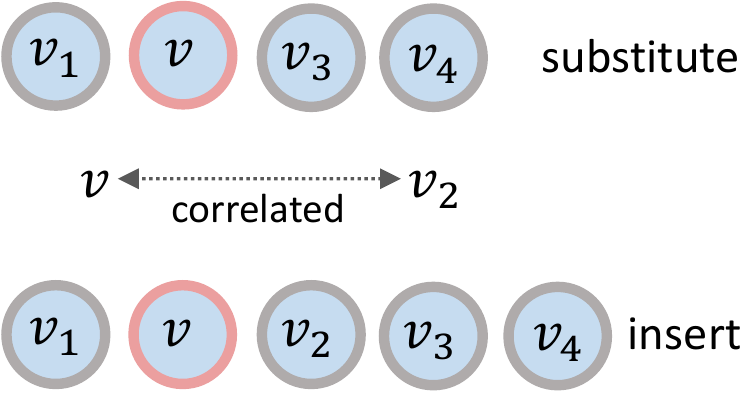}
    \caption{Informative}
    \label{fig:informative_aug}
\end{subfigure}
\caption{Five augmentations adopted in this paper. The original sequence is $[v_1,v_2,v_3,v_4]$. 
Informative augmentations incorporate correlations
among items. }
\label{fig:aug_methods}
\end{figure}

We first review and formulate existing random augmentation methods~\cite{wu2020clear,xie2020contrastive,zhou2020s3}, and then introduce two novel informative augmentation methods. We also describe how to augment sequences based on their length. This section assumes the original sequence being $s_{u}=[v_{1}, v_{2}, \dots, v_{n}]$. 
Toy examples of those methods are illustrated in Figure~\ref{fig:aug_methods}, where $n=4$.

\subsubsection{\textbf{Random Augmentation}}
Three random operators 
introduced by~\cite{zhou2020s3,xie2020contrastive} are included in this paper to create various views of original sequences:
\begin{itemize}
    \item \textbf{Crop ($\mathsf{C}$).} Randomly select a continuous sub-sequence of the original
    sequence starting from position $i$:
    \begin{equation}
        s_{u}^{\mathsf{C}}=\mathsf{C}(s_{u})=[v_{i}, v_{i+1}, ...,v_{i+c-1}], 
    \end{equation}  
    where $c=\lceil \eta n \rceil$ is the sub-sequence length that is controlled by a hyperparameter $\eta$ and $0\leq\eta\leq 1$. $\lceil \cdot \rceil$ is the ceiling function. 
    \item \textbf{Mask ($\mathsf{M}$).} Randomly mask $l=\lceil\mu n\rceil$ items in a sequence:
    \begin{equation}
        s_{u}^{\mathcal{M}}=\mathsf{M}(s_{u})=[v_{1}', v_{2}', ...,v_{n}'],
    \end{equation} 
   where $v_{i}'$ is the `mask' if $v_{i}$ is a selected item, otherwise $v_{i}'=v_{i}$. $l$ is controlled by a hyper-parameter $\mu$, where $0\leq\mu\leq 1$.
   
    \item \textbf{Reorder ($\mathsf{R}$).} Randomly shuffle a sub-sequence 
    $\left[v_{i},\cdots, v_{i+r-1}\right]$ 
    of $s_{u}$ as $[{v}'_{i}, \ldots, {v}'_{i+r-1}]$:
    \begin{equation}
        s_{u}^{\mathsf{R}}=\mathsf{R}(s_{u})=[v_{1}, v_{2},\cdots, {v}'_{i}, \cdots, {v}'_{i+r-1}, \cdots, v_{n}],
    \end{equation} 
    where $r=\lceil\omega n\rceil$ is the 
    sub-sequence length and $0\leq\omega\leq 1$.
\end{itemize}

In general, good augmentations for one instance should give views sharing task-oriented information~\cite{tian2020makes}, thus being positive pairs. 
In our case, views from one sequence are supposed to maintain the original sequential correlations, which otherwise could yield less confident positive pairs.  
However, those random augmentations break the item correlations in sequences, especially for short sequences. 
For example, reordering a sequence leads to changed item correlations in the sub-sequence. 
This impact is 
enlarged when fewer items are 
present in sequences.

Moreover, random augmentations   exaggerate cold-start issues in sequences, \textit{e.g.} masking a short sequence induces few items in a sequence. As such,
the sequence encoder may {fail to learn high quality representations}, 
and thus fail
to characterize item relationships in sequences.
Contrasting on low quality sequence representations further deteriorates the robustness of SSL. 

Next, we will introduce two novel informative augmentation methods, which advance random augmentations regarding the mentioned concerns. 

\subsubsection{\textbf{Informative Augmentation}} 
\label{sec:inform-augmentation-operators}
We propose two informative augmentation operators leveraging item correlations
to generate robust augmented sequences, 
as
illustrated in Figure~\ref{fig:informative_aug}.
\begin{itemize}
    \item \textbf{Substitute ($\mathsf{S}$).} The substitution augmentation is motivated by the real case~\cite{chen2020try} where recommending substitutable items to users 
    expands the chance of discovering their actual interests.
    Correspondingly, substituting items in sequences with highly correlated item injects less corruption to the original sequential information, which thus yields confident positive pairs of views.   
    Formally, 
    we randomly select 
    $k$ different
    indices
    $\{\mathrm{idx}_{1}, \mathrm{idx}_{2},
    \dots, \mathrm{idx}_{k}\}$ in the sequence $s_u$, where $k=\lceil\alpha n\rceil$ and $\mathrm{idx}_{i}\in[1,2,\dots,n]$. $\alpha\in[0,1]$ is substitution ratio.
    Then we replace each with an correlated item based on the selected 
    indices.
    The sequence after substitution will be:
    \begin{equation}
        s_{u}^{\mathsf{S}}=\mathsf{S}(s_{u})=[v_{1}, v_{2}, ...,\bar{v}_{\mathrm{idx}_i}, \dots, v_{|s_{u}|}], 
    \end{equation}   
    where $\bar{v}_{\mathrm{idx}_i}$ is a correlated item with $v_{\mathrm{idx}_i}$.  
    We will describe how to choose the correlated items in Section~\ref{sec:item-correlations}.

    \item \textbf{Insert ($\mathsf{I}$).} 
    In practice, recorded interactions are less than the 
    complete consumption
    behavior of users, \textit{e.g.} 
    transactions from other sources are missing~\cite{liu2019jscn}.
    Consequently, training sequences may fail to track comprehensive user dynamics
    and item correlations. 
    To this end, we construct augmentations by
    inserting items into sequences, 
    in order to
    `complete' the sequence.
    The number of inserted items is controlled 
    by the ratio $\beta\in[0,1]$. 
    We first randomly select $k$ different
    indices
    $\{\mathrm{idx}_{1}, \mathrm{idx}_{2},\dots, \mathrm{idx}_{k}\}$ in the sequence $s_{u}$,
    where $k=\lceil\beta n\rceil$ and $\mathrm{idx}_{i}\in[1,2,\dots,n]$. 
    We
    then insert 
    correlated items at those indices.
    After insertion,
    the sequence is:
    \begin{equation}
        s_{u}^{\mathsf{I}}=\mathsf{I}(s_{u})=[v_{1}, v_{2}, ...,\bar{v}_{\mathrm{idx}_i},v_{\mathrm{idx}_i}, \dots, v_{n}], 
    \end{equation}
    where $i \in \{1,2,\dots,k\}$, and $\bar{v}_{\mathrm{idx}_i}$ is
    the most correlated item with $v_{\mathrm{idx}_i}$. The length of the augmented sequence $s_{u}^{\mathsf{I}}$ is $k+n$.
\end{itemize}
Both augmentations are informative as they perturb the sequence with correlated items. As such, augmented sequences 
have higher confidence of being
positive pairs. Moreover, adding correlated items to sequences extends the number of user interactions, which alleviates user cold-start issues. Next, we introduce how to choose correlated items for informative augmentations.

\subsubsection{\textbf{Item Correlations}}\label{sec:item-correlations}
Informative augmentations require 
inferring correlations among items
to
perturb sequences. Thus, it is necessary to design a simple yet effective method for calculating item correlations. We first introduce two straightforward methods and then demonstrate a  hybrid method adopted in this work. 

The first 
is \textit{memory-based} correlation. Collaborative signals are crucial in recommender systems.
Therefore, we employ item-based collaborative filtering 
considering inverse user frequency (ItemCF-IUF~\cite{breese2013empirical}) to measure item correlations 
due to its simplicity and effectiveness.
Formally, the memory-based correlation score between item $i$ and $j$ is defined as:
\begin{equation}
    \mathrm{{Cor}_{o}}(i, j)=\frac{1}{\sqrt{|\mathcal{N}(i)||\mathcal{N}(j)|}}\sum_{u\in \mathcal{N}(i)\cap \mathcal{N}(j)}\frac{1}{\log(1+|\mathcal{N}(u)|)},
\end{equation}
where $u$ is a user, $|\mathcal{N}(i)|$, and $|\mathcal{N}(j)|$ are the number of users who have interacted with $i$ and $j$, respectively. 
Memory-based correlation 
treats
two items as being more correlated if they share a higher ratio of common users.

The other 
is \textit{model-based} correlation. This correlation directly infers correlations by measuring the similarity between item representations.
Since item representations are jointly learned with the encoder, 
this method is thus model-based. 
In this work, we adopt the dot-product as the similarity metric.
Given the representations of items $i$ and $j$ as $\mathbf{e}_{i}$ and $\mathbf{e}_{i}$, the model-based correlation score is defined as:
\begin{equation}
    \mathrm{Cor_{e}}(i, j)=\mathbf{e}_{i}\cdot \mathbf{e}_{j}.
\end{equation}

In this paper, we fuse 
the \textit{memory-based} and \textit{model-based} correlations 
as a \textit{hybrid} correlation $\mathrm{Cor_{h}}$, which is defined as the highest value between memory-based and model-based correlations as:
\begin{equation}
    \mathrm{Cor_{h}}(i,j) = \max ( \mathrm{\overline{Cor}_{o}}(i, j),\mathrm{\overline{Cor}_{e}}(i, j) ),
\end{equation}
where $\mathrm{\overline{Cor}_{o}}$ and $\mathrm{\overline{Cor}_{e}}$ are the normalized score of memory-based and model-based correlations, respectively. We normalize (min-max normalization) them to be comparable.
Additionally, since item representations learned at early epochs are not informative, 
we initially use $\mathrm{Cor_{o}}$ to infer correlated items. Then, after training with $E$ epochs, we switch to $\mathrm{Cor_{h}}$, where $E$ is a hyper-parameter. 


\subsubsection{\textbf{Augment w.r.t. Sequence Length}}
\label{sec:short-long-sequences}
The length of interaction
sequences
tends to
follow a long-tail distribution~\cite{liu2021augmenting},
in which most 
sequences are short. Considering that
short sequences are more sensitive to 
random item perturbations, 
we should 
carefully
choose 
augmentation operations on short sequences.  
In this work, we employ different augmentation operators sets to 
sequences with respect to their lengths. We use a hyper-parameter $K$ to decide whether a sequence is 
short or long,
and then apply data augmentation as follows:
\begin{equation}\label{eq:aug_length}
s_{u}^{{a}} =
\left\{
             \begin{array}{lr}
             {{a}}(s_{u}), ~{{a}} \sim\{\mathsf{S}, \mathsf{I}, \mathsf{M}\},&  |s_{u}|\leq K  \\
             {{a}}(s_{u}), ~{a} \sim\{\mathsf{S}, \mathsf{I}, \mathsf{M}, \mathsf{C}, \mathsf{R}\}, & |s_{u}| > K
             \end{array}
\right.
\end{equation}    
where ${a}$ is an
augmentation operator selecting 
from the corresponding augmentation set.
Though $\mathsf{M}$ is a random augmentation, we also include it in the augmentation set for short sequences. 
The reason are twofold: Firstly,  mask augmentation implicitly models item relationships in sequence, which is similar to the next-item prediction target. 
Secondly, masking operation encourages items near masked items being closer, which models high-order item relationships in the original sequence.

\subsection{Contrastive Self-Supervision}
Contrastive SSL optimizes encoders by
maximizing the agreements between
`positive' pairs, which are two augmentations from one sequence.
To be specific, given a minibatch of sequences $\{s_{u}\}_{u=1}^{N}$, 
we sample two augmentation operators for each sequence $s_u$, 
which obtains $2N$ augmented sequences as:
\begin{equation}
    \{\tilde{s}_{1}, \tilde{s}_{2}, \dots, \tilde{s}_{2u-1}, \tilde{s}_{2u}, \dots, \tilde{s}_{2N-1}, \tilde{s}_{2N}\},
\end{equation}
where $ u \in \{1,2, \dots, N\}$.
Following~\cite{chen2020simple}, 
each pair $(\tilde{s}_{2u-1}, \tilde{s}_{2u})$ is treated as a positive pair, 
and the other $2(N-1)$ augmented views are considered as negative samples for this pair. The augmented sequences are then encoded as in Eq.~(\ref{eq:seq_encoder})
with a shared sequence encoder. 
For each sequence pair $(\tilde{s}_{2u-1}, \tilde{s}_{2u})$,
their representations are $(\tilde{\mathbf{h}}_{2u-1}, \tilde{\mathbf{h}}_{2u})$.  We adopt
the NT-Xent loss~\cite{chen2020simple} for optimization as follows:
\begin{equation}
\mathcal{L}_{\mathrm{ssl}}(\mathbf{\Tilde{h}}_{2u-1}, 
                    \mathbf{\tilde{h}}_{2u}) = 
        - \log \frac{\exp(\text{sim}(\mathbf{\tilde{h}}_{2u-1},                       \mathbf{\tilde{h}}_{2u}))}
        {\sum_{m=1}^{2N}\mathbbm{1}_{m\neq 2u-1}\exp(\text{sim}(\mathbf{\tilde{h}}_{2u-1}, \mathbf{\tilde{h}}_{m}))},
\end{equation}  
where $\mathrm{sim}(\cdot)$ is a  dot product to measure the similarity between two augmented views, 
and $\mathbbm{1}_{|m\neq 2u-1|}\in \{0, 1\}$ is an indicator function.
Since the output embeddings from a Transformer encoder are position-wise, 
we thus concatenate representations at all positions
as the sequence representation. We illustrate the contrastive SSL framework in Figure~\ref{fig:framework}(c). 

\subsection{Multi-Task Training}
Because 
next-item prediction and contrastive SSL both 
model item relationships in sequences, to boost 
sequential recommendation performance with the contrastive SSL objective,
we leverage a multi-task strategy to optimize 
them jointly as follows:

\begin{algorithm}[htb]
\SetKwInput{KwInput}{Input}                
\SetKwInput{KwOutput}{Output} 
\KwInput{hyparameters $\alpha$, $\beta$, $K$, $E$, and $\lambda$, max training epochs $B$, batch size $N$, and a memory-based correlation matrix $\mathrm{Cor_{o}}$.}
\KwOutput{$\mathsf{SeqEnc}(\cdot)$.}
 \While{ $epoch \leq B$}{
    \tcp{Item correlation selection (Sec.~\ref{sec:item-correlations})}
    
    \eIf{$\mathit{epoch} \leq E$}{
    Leverage $\mathrm{Cor_{o}}$ for informative augmentations.
    }{
    Leverage $\mathrm{Cor_{h}}$ for informative augmentations.
    }
    \tcp{End-to-end optimization}
    \For{sampled minibatch $\{s_{u}\}_{u=1}^{N}$}{
        \For{$u \in \{1, 2, \cdots, N\}$}{
              \tcp{Construct two augmentations.}
              \eIf{$|s_{u}| \leq K$}{
               sample two operators: $a_{1}^{u}, a_{2}^{u} \sim \{\mathsf{I}, \mathsf{S}, \mathsf{M}\}$
               }{
               sample two operators: $a_{1}^{u}, a_{2}^{u} \sim \{\mathsf{I}, \mathsf{S}, \mathsf{M}, \mathsf{C}, \mathsf{R}\}$
              }
              $\tilde{s}_{2u-1}=a_{1}^{u}(s_{u}), \tilde{s}_{2u}=a_{2}^{u}(s_{u})$
              
              \tcp{Representation learning via $\mathsf{SeqEnc}$} 
              
              $\tilde{\mathbf{h}}_{u} = \mathsf{SeqEnc}(s_{u})$ \\
              $\tilde{\mathbf{h}}_{2u-1},\tilde{\mathbf{h}}_{2u} = \mathsf{SeqEnc}(\tilde{s}_{2u-1}),\mathsf{SeqEnc}(\tilde{s}_{2u})$
        }
         
        \tcp{ Multi-Task optimization}
          $\mathcal{L} = \frac{1}{N}\sum_{u=1}^{N}\mathcal{L}_{\mathrm{rec}}(\tilde{\mathbf{h}}_{u}) + \lambda \frac{1}{2N}\sum_{u=1}^{N}[\mathcal{L}_{\mathrm{ssl}}(\tilde{\mathbf{h}}_{2u-1}, \tilde{\mathbf{h}}_{2u})+
          \mathcal{L}_{\mathit{ssl}}(\tilde{\mathbf{h}}_{2u}, \tilde{\mathbf{h}}_{2u-1})]$ \\
          
          update network $\mathsf{SeqEnc}(\cdot)$ to minimize $\mathcal{L}$
    }
    
 }

 \caption{\modelname training.}
 \label{alg:coserec}
\end{algorithm}

\begin{equation}
\mathcal{L} = \mathcal{L}_{\mathrm{rec}} + \lambda \mathcal{L}_{\mathrm{ssl}},
\end{equation}
where $\lambda$ is a hyper-parameter to control the intensity of contrastive SSL. 
Algorithm~\ref{alg:coserec} summarizes the training process of \modelname.
Alternatively, a two-stage optimization, \textit{i.e.} first pre-training the encoder with SSL and then fine-tuning with next-item prediction,
can be adopted. We provide comparisons in Section~\ref{sec:pre-train-effect-and-insert-benefits}, which indicates joint training is better than two-stage training.

\section{Experiments}
In this section, we conduct experiments 
on three public datasets 
and answer the following Research Questions (\textbf{RQs}):

\begin{itemize}
    \item \textbf{RQ1:} How does 
    \modelname perform on sequential recommendations compared with existing methods? 
    \item \textbf{RQ2:} What are the optimal augmentation methods for sequential recommendation?
    \item \textbf{RQ3}: Can \modelname perform robustly against data-sparsity and noisy interactions issues? 
    \item \textbf{RQ4}: How do different settings influence
    \modelname's performance?
\end{itemize}

\subsection{Experimental Setting}

\subsubsection{\textbf{Datasets}}
We conduct experiments on three public datasets collected from two real-word platforms. \emph{Beauty} and \emph{Sports} 
are two subcategories of Amazon review data
introduced in~\cite{mcauley2015image}. The Yelp\footnote{https://www.yelp.com/dataset} dataset is a dataset for business recommendation.

We follow common 
practice
in~\cite{zhou2020s3,xie2020contrastive} to preprocess the datasets. 
The numeric ratings or 
the presence of a review are treated as positive 
instances while others as negative instances. 
We only keep the `5-core' datasets, 
i.e. every user bought in total at least 5 items
and vice versa each item was bought by at least
5 users.  Table~\ref{tab:dataset-information} shows the statistics of the datasets after preprocessing.

\begin{table}[htb]
  \caption{Dataset information.}
  \label{tab:dataset-information}
  \setlength{\tabcolsep}{1.5mm}{
  \begin{tabular}{lccccc}
    \toprule 
    Dataset & \# Users & \# Items & \# Actions & Avg. length & Sparsity \\
    \midrule
    Beauty      & 22,363   & 12,101   & 198,502    & 8.9         & 99.73\% \\
    Sports      & 35,598   & 18,357   & 296,337    & 8.3         & 99.95\% \\
    Yelp        & 30,431   & 20,033   & 316,354     & 10.3        & 99.95\% \\
  \bottomrule
\end{tabular}}
\end{table}

\subsubsection{\textbf{Evaluation Metrics}} We follow ~\cite{kang2018self,sun2019bert4rec} to evaluate the model as next-item prediction. 
We rank the prediction on whole item set without negative sampling, which otherwise leads biased discoveries~\cite{krichene2020sampled}. 
Performance is
evaluated on 
top-$k$ ranking metrics including \textit{Hit Ratio}$@k$ ($\mathrm{HR}@k$), and \textit{Normalized Discounted 
Cumulative Gain}$@k$ ($\mathrm{NDCG}@k$). We report $\mathrm{HR}$ and $\mathrm{NDCG}$ with $k\in\{5, 10, 20\}$.

\subsubsection{\textbf{Baseline Methods}}
We include three groups of 
baseline methods for comparison. 

\begin{itemize}
    \item \textit{Non-Sequential models:}
    PopRec 
    is based on the popularity of items;
    BPR-MF~\cite{rendle2012bpr} 
    is a matrix factorization model with 
    a
    pairwise 
    Bayesian Personalized Ranking loss.
    \item \textit{Transformer-based SR models:} SASRec~\cite{kang2018self} is one of the state-of-the-art Transformer-based
    baselines for SR,
    Bert4Rec~\cite{sun2019bert4rec} 
    extends SASRec with
    bidirectional self-attention modules.
    S$^3$\text{-Rec}~\cite{zhou2020s3} uses self-supervised
    learning to capture relationships among item and 
    associated attributes.
    We remove its modules for mining on attributes as we have no attributes for items,
    namely S$^3\text{-Rec}_{m}$.
    CL4SRec~\cite{xie2020contrastive} fuses
    contrastive SSL with 
    Transformer-based SR model. It only has random augmentation methods for SSL.
    \item \textit{Other SR models:} GRU4Rec~\cite{hidasi2015session} 
    is an RNN-based approach, and Caser~\cite{tang2018personalized} is a CNN-based model.
    
\end{itemize}

\subsubsection{\textbf{Implementation Details}}

\begin{table*}[htb]
  \caption{Performance comparisons of different methods. 
  The best score is bolded in each row, and the second best is underlined.
  The last two columns are the relative
  improvements compared with SASRec and the best baseline results.}
  \label{tab:main-results}
  \setlength{\tabcolsep}{1.3mm}{
  \begin{tabular}{ll|cc|cc|cccc|c|c|c}
    \toprule
    \multirow{2}{*}{Dataset} & \multirow{2}{*}{Metric} &
    \multirow{2}{*}{PopRec} & \multirow{2}{*}{BPR} &
    \multirow{2}{*}{GRU4Rec} & \multirow{2}{*}{Caser} &
    \multirow{2}{*}{SASRec} & \multirow{2}{*}{BERT4Rec} &
    \multirow{2}{*}{S$^3\text{-Rec}_{m}$} & \multirow{2}{*}{CL4SRec} &
    \multirow{2}{*}{CoSeRec} & \multicolumn{2}{c}{Improv. v.s.} \\
    \cline{12-13}
    & & & & & & & & & & & SASRec & All \\
    

    \hline
    \midrule
    \multirow{6}{*}{Beauty}  & HR@5   & 0.0080 & 0.0212  & 0.0111 & 0.0251 & 0.0374 & 0.0351 & 0.0189 & \underline{0.0401} & \textbf{0.0537} & 43.58\% & 33.92\%\\
                             & HR@10  & 0.0152 & 0.0372& 0.0162 & 0.0342 & 0.0575 & 0.0601 & 0.0307 & \underline{0.0642} & \textbf{0.0752} & 30.78\% & 17.13\%\\
                             & HR@20  & 0.0217 & 0.0589 & 0.0478 & 0.0643 & 0.0901 & 0.0942 & 0.0487 & \underline{0.0974} & \textbf{0.1041} & 15.54\% & 6.88\%\\
                             & NDCG@5 & 0.0044 & 0.0130 & 0.0058 & 0.0145 & 0.0241 & 0.0219 & 0.0115 & \underline{0.0268} & \textbf{0.0361} & 49.79\% & 34.7\%\\
                             & NDCG@10& 0.0068 & 0.0181 & 0.0075 & 0.0226 & 0.0305 & 0.0300 & 0.0153 & \underline{0.0345} & \textbf{0.0430} & 40.98\% & 24.64\%\\
                             & NDCG@20& 0.0084 & 0.0236 & 0.0104 & 0.0298 & 0.0387 & 0.0386 & 0.0198 & \underline{0.0428} & \textbf{0.0503} & 29.97\% & 17.52\%\\
    \midrule
    \multirow{6}{*}{Sports}  & HR@5    & 0.0056 & 0.0141  & 0.0162 & 0.0154 & 0.0206 & 0.0217 & 0.0121 & \underline{0.0231} & \textbf{0.0287} & 39.32\% & 24.24\%\\
                             & HR@10   & 0.0094 & 0.0216 & 0.0258 & 0.0261 & 0.0320 & 0.0359 & 0.0205 & \underline{0.0369} & \textbf{0.0437} & 36.56\% & 18.43\%\\
                             & HR@20   & 0.0192 & 0.0323 & 0.0421 & 0.0399 & 0.0497 & \underline{0.0604} & 0.0344 & 0.0557 & \textbf{0.0635} & 27.77\% & 8.60\%\\
                             & NDCG@5  & 0.0041 & 0.0091 & 0.0103 & 0.0114 & 0.0135 & 0.0143 & 0.0084 & \underline{0.0146} & \textbf{0.0196} & 45.19\% & 34.25\%\\
                             & NDCG@10 & 0.0053 & 0.0115 & 0.0142 & 0.0135 & 0.0172 & 0.0190 & 0.0111 & \underline{0.0191} & \textbf{0.0242} & 40.7\% & 26.7\%\\
                             & NDCG@20 & 0.0078 & 0.0142 & 0.0186 & 0.0178 & 0.0216 & \underline{0.0251} & 0.0146 & 0.0238 & \textbf{0.0292} & 35.19\% & 15.14\%\\
    \midrule
    \multirow{6}{*}{Yelp}  & HR@5     & 0.0057 & 0.0127 & 0.0152 & 0.0142 & 0.0160 & 0.0196 & 0.0101 & \underline{0.0227} & \textbf{0.0241} & 50.63\% & 6.17\%\\
                             & HR@10  & 0.0099 & 0.0216 & 0.0248 & 0.0254 & 0.0260 & 0.0339 & 0.0176 & \underline{0.0384} & \textbf{0.0395} & 51.92\% & 2.86\%\\
                             & HR@20  & 0.0164 & 0.0346 & 0.0371 & 0.0406 & 0.0443 & 0.0564 & 0.0314 & \underline{0.0623} & \textbf{0.0649} & 46.5\% & 4.17\%\\
                             & NDCG@5 & 0.0037 & 0.0082 & 0.0091 & 0.008 & 0.0101 & 0.0121 & 0.0068 & \underline{0.0143} & \textbf{0.0151} & 49.5\% & 5.59\%\\
                             & NDCG@10& 0.0051 & 0.0111 & 0.0124 & 0.0113 & 0.0133 & 0.0167 & 0.0092 & \underline{0.0194} & \textbf{0.0205} & 54.14\% & 5.67\%\\
                             & NDCG@20& 0.0067 & 0.0143 & 0.0145 & 0.0156 & 0.0179 & 0.0223 & 0.0127 & \underline{0.0254} & \textbf{0.0263} & 46.93\% & 3.54\%\\
                             
  \bottomrule
\end{tabular}}
\end{table*}

We use 
code provided by
the authors for
S3-Rec\footnote{https://github.com/RUCAIBox/CIKM2020-S3Rec}, Bert4Rec\footnote{https://github.com/FeiSun/BERT4Rec} and
Caser\footnote{https://github.com/graytowne/caser\_pytorch}
methods.
We implement GRU4Rec\footnote{{https://github.com/slientGe/Sequential\_Recommendation\_Tensorflow}} and BPRMF\footnote{https://github.com/xiangwang1223/neural\_graph\_collaborative\_filtering} methods
based on public resources.
We tune the mask proportion of
BERT4Recc with $\{0.2, 0.4, 0.6, 0.8\}$.
We set all other 
hyper-parameters of each method as reported in the papers. 

We implement our method in PyTorch. For common hyparparameters, we follow~\cite{xie2020contrastive} to set the number of the 
self-attention blocks and attention heads as $2$,
the embedding dimension as 64, and the maximum sequence length as $50$.
We tune $\alpha$, $\beta$, $K$, $E$, and
$\lambda$ within the ranges of $[0.1,0.9]$, 
$[0.1,0.9]$, 
$\{4, 12, 20\}$, $\{0, 20, 40, 60, 80, 100\}$, 
and 
$\{0.1, 0.2, 0.3, 0.4, 0.5\}$ respectively. 
For hyperparameters in 
random augmentation operators 
$\eta$, $\mu$, and $\omega$, we set the optimal
values reported by~\cite{xie2020contrastive}.
We adopt early stopping 
on the validation set 
if the performance does not improve for 40 epochs,
and report results on test set.
The model is optimized by Adam optimizer~\cite{kingma2014adam} with a
learning rate of 0.001, $\beta_{1}=0.9$, $\beta_{2}=0.999$, and
batch size of 256.

\subsection{Overall Performance Comparison (RQ1)}

Table~\ref{tab:main-results} shows the performance
of all methods on three datasets. 
We 
make the
follow observations:

\begin{itemize}
    \item \textit{Overall Comparison. }
    (1) Non-sequential models perform
    worse than sequential recommendation methods.
    This 
    suggests
    the importance of mining sequential patterns
    for next-item prediction. 
    For sequential models, Transformer-based methods
    achieve better performance than other type of SR models,
    which indicates that self-attention mechanisms are 
    more
    effective at capturing sequential patterns 
    than CNN and RNNs.  
    (2) S$^{3}\mathit{Rec}_{m}$, 
    though leveraging
    self-supervision signals by masking
    items
    in sequences,
    performs worse than SASRec. We hypothesize the poor performance results from two reasons: the two-stage training preventing information sharing between SSL and next-item prediction targets, and the weak supervision signals without informative augmentations. 
    (3) CL4SRec consistently performs better than other baselines, which verifies the efficacy of employing contrastive SSL in sequential recommendation. However, it still performs worse than \modelname, since it has no informative augmentations.
    (4) The proposed \modelname consistently
    outperforms other models on all datasets in all evaluation metrics.
    The improvements comparing with the best baseline
    range from 3.54\% to 30.88\% in terms of $\mathrm{HR}$ and $\mathrm{NDCG}$.
    Different from baselines, our proposed method leverages
    contrastive SSL signals with informative augmentation methods, incorporating length skewness of sequences,
    and optimize the model with multi-task strategy. 
    \item \textit{Comparison with SASRec and CL4SRec.}
    (1)
    We observe that compared with SASRec, the performance of \modelname 
    is significantly better,
    ranging from $14.54\%$ to 
    $54.14\%$ in $\mathrm{HR}$ and $\mathrm{NDCG}$ on three datasets. Since both \modelname and SASRec adopt the same sequence encoder, those results
    demonstrate the necessity of 
    leveraging the constrastive SSL signals for SR. 
    (2) Compared with CL4SRec which employs random augmentations
    for contrastive SSL task,
    \modelname acheives 22.47\%, 23.39\% and 4.67\% average relative improvement on Beauty, Sports, and Yelp datasets respectively.
    The performance gains result from the informative augmentation, which is first proposed in this work.
    (3) We also observe larger average improvements on Sports and Beauty
    compared with Yelp. We hypothesis the difference is because the average sequence
    length of Sports and Beauty is shorter than that of Yelp, thus
    suffering severer
    `cold-start' issues. Therefore,
    SSL contributes more benefits 
    to
    the sequence encoding on the Sports and Beauty datasets. 
\end{itemize}

\subsection{\textbf{Augmentation Analysis (RQ2)}} 

In this paper, we include five augmentation operators for the contrastive SSL task.
To study impacts of these
augmentation methods and find the optimal choice, we conduct three groups of ablation studies on
the Sports and Beauty datasets. 
The first is a `leave-one-out' setting in which only one augmentation
operator is removed from the augmentation set $\{\mathsf{M}, \mathsf{C}, \mathsf{R}, \mathsf{S}, \mathsf{I}\}$ at a time. 
The second
is `pair-wise' 
where we only apply two augmentations to create
two types of views for contrastive SSL tasks in each model. 
Lastly, we investigate the optimal augmentations
for short sequences.

\subsubsection{\textbf{Leave-One-Out Comparison}}

\begin{figure}[htb]
  \centering
  \includegraphics[width=\linewidth]{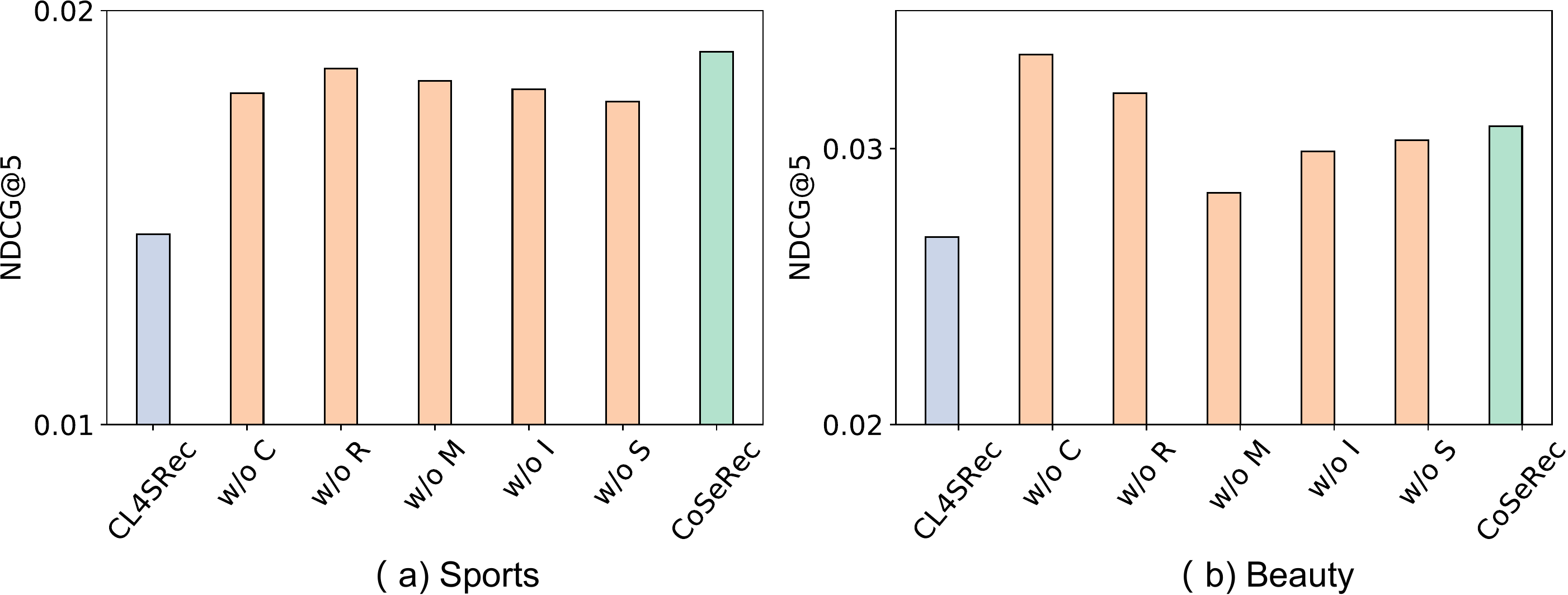}
  \caption{Performance comparison (in NDCG@5) w.r.t.~different augmentation sets 
  on Sports and Beauty datasets. ‘w/o' indicates one type is removed from the full augmentation set $\{\mathsf{M}, \mathsf{C}, \mathsf{R}, \mathsf{S}, \mathsf{I}\}$. \modelname adopts the full set.}
  \label{fig:ablation-aug}
\end{figure}

Figure~\ref{fig:ablation-aug} shows the comparison results 
under the `leave-one-out' setting. We included 
CL4SRec for comparison since
it utilizes random augmentations $\{\mathsf{M}, \mathsf{C}, \mathsf{R}\}$.
We observe that
without 
the
`Insert' or `Substitute' operator, the performance consistently decreases on both 
datasets.
Without random operators such as `Crop' or `Reorder', 
performance even significantly 
\textit{increases} on 
the
Beauty dataset. This phenomenon implies that 
existing random augmentations, such as
randomly cropping or reordering a sequence without considering item correlations, demolish item relationships in sequences, which leads to less confident positive pairs and thus impairs
the efficacy of the contrastive SSL learning. 

\subsubsection{\textbf{Pair-Wise Comparison}}

\begin{figure}[htb]
  \centering
  \includegraphics[width=\linewidth]{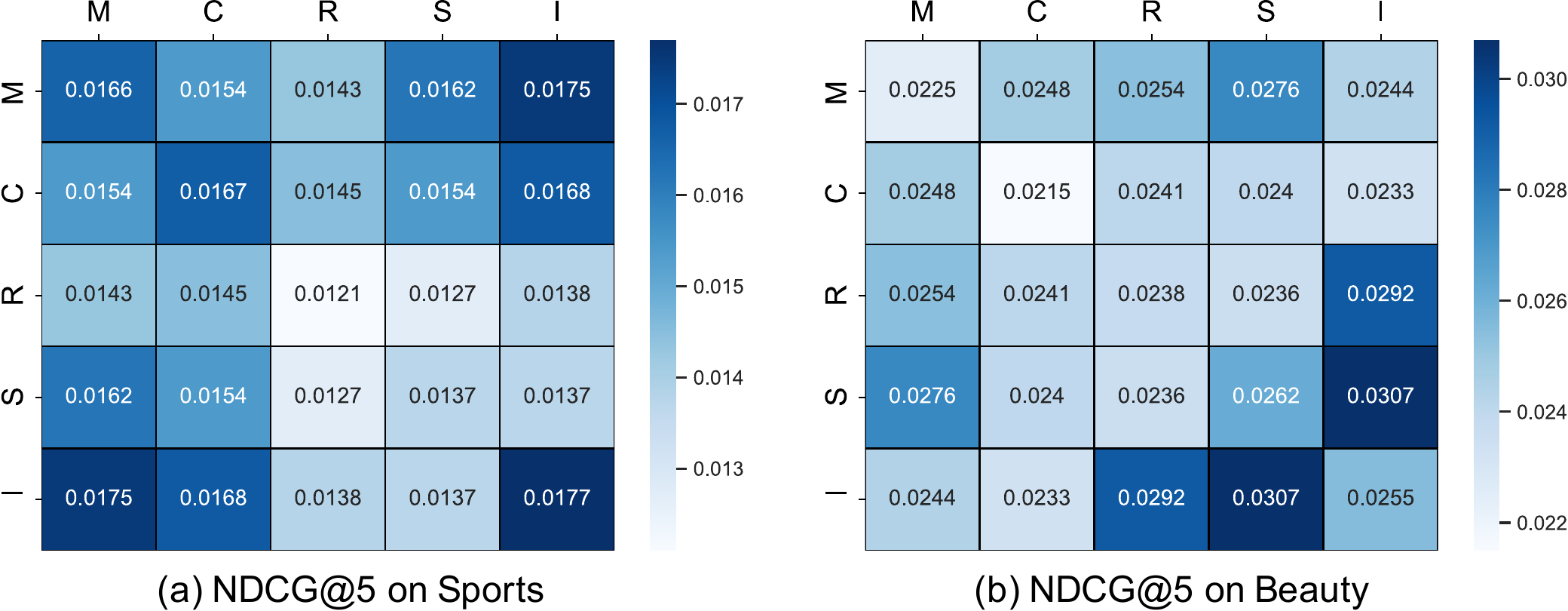}
  \caption{Performance comparison (in $\mathrm{NDCG}@5$) w.r.t.~different augmentation pairs on Sports and Beauty datasets. }
  \label{fig:ablation-pair-wise}
\end{figure}

Figure~\ref{fig:ablation-pair-wise} shows the comparison results 
on the Sports and Beauty datasets 
under the `pair-wise' setting.
We 
observe that: 
(1) For pair of the same operator (diagonal values), 
the best performance is $\mathsf{I}$ and $\mathsf{S}$ on Sports and Beauty dataset, respectively.
This demonstrates 
our proposed informative operators are better than those random augmentations
for contrastive SSL. 
(2) A pair of 
different operators 
performs better than 
two which are the same
in general. 
For example in Beauty, $(\mathsf{I},\mathsf{S})$ pair achieves
0.0307 while $(\mathsf{I},\mathsf{I})$ and $(\mathsf{S},\mathsf{S})$ pairs achieves 0.0255 and 0.0262 in $\mathrm{NDCG}@5$ respectively. 
Comparing
with Figure~\ref{fig:ablation-aug}(b), 
we observe that having multiple augmentation operators
perform better than 
one or two augmentation operators. 
For example, with $\{\mathsf{M}, \mathsf{R}, \mathsf{S}, \mathsf{I}\}$ augmentation set,
the model achieves 0.0334 in $\mathrm{NDCG}@5$ on Beauty which is higher than all 
pair-wise based model 
results.
These observations imply that having multiple views for contrastive SSL is beneficial 
as it allows 
the
model to capture more mutual information from different views.
(3) `$\mathsf{S}$'
performs
well on Beauty but not on Sports under the pair-wise setting. 
However, in Figure~\ref{fig:ablation-aug}(a), 
model
without `S' augmentation has the largest performance drop
compared to others.
This indicates that the `S' operator is more suitable 
for providing a complementary view for 
contrastive
SSL learning.
\begin{figure}[htb]
  \centering
  \includegraphics[width=\linewidth]{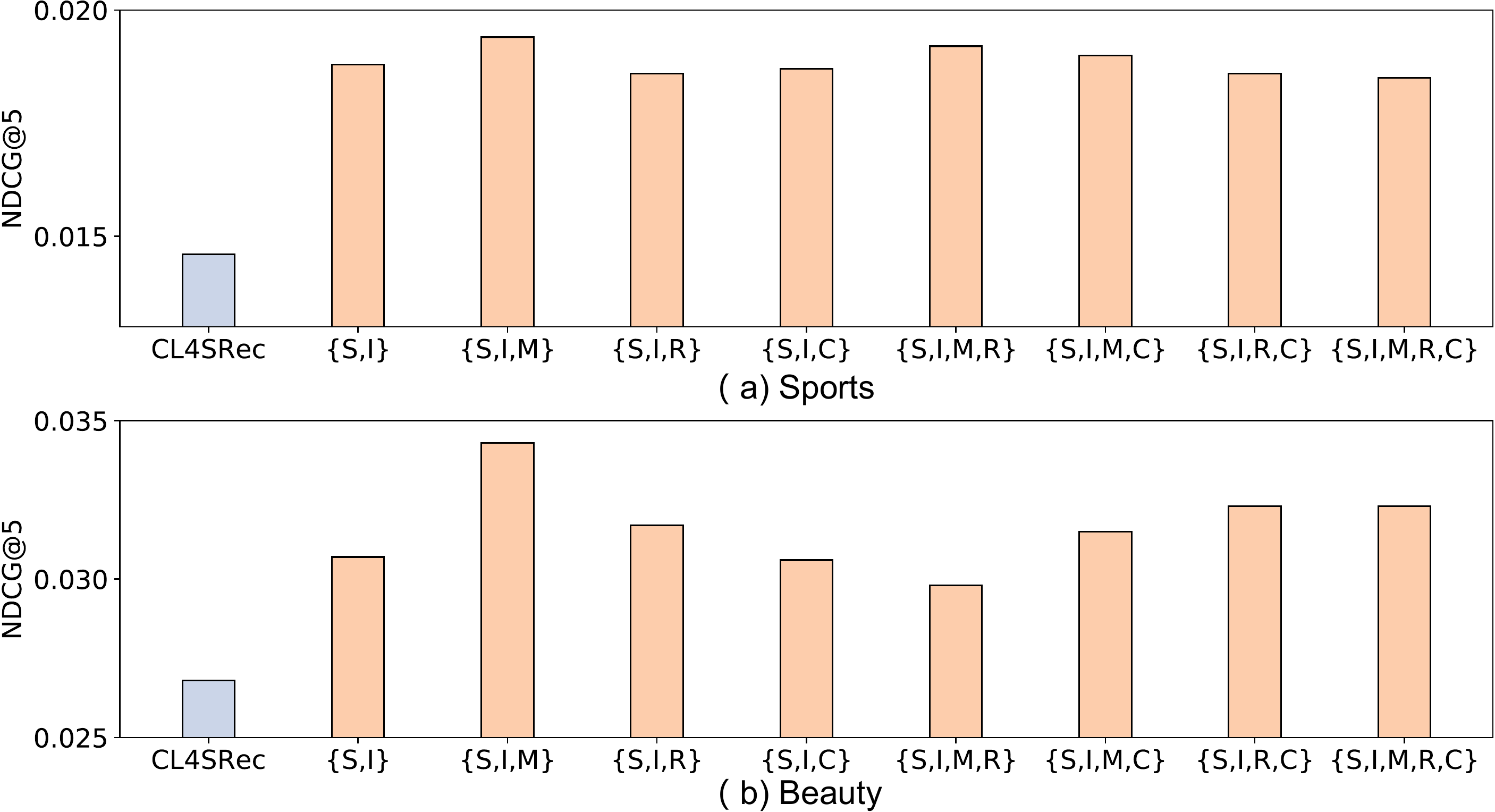}
  \caption{Performance comparison w.r.t. different augmentation sets for short sequences on Sports and Beauty datasets.}
  \label{fig:ablation-short-seq-aug}
\end{figure}

\subsubsection{\textbf{Augmentation Set for Short Sequences}}
As described in Sec.~\ref{sec:short-long-sequences},
short sequences are more sensitive to random augmentations, for which we only augmented by `$\mathsf{M}$', `$\mathsf{S}$', and `$\mathsf{I}$' augmentation operators. In this section, we explore
the optimal augmentation combinations for short sequences to verify our claims. 

Figure~\ref{fig:ablation-short-seq-aug} shows the 
performance
with different combinations of augmentations
for short sequences. For example, `$\{\mathsf{S},\mathsf{I},\mathsf{M}\}$' denotes adopting  all `$\mathsf{S}$', `$\mathsf{I}$', and `$\mathsf{M}$' augmentations for short sequences.  We observe that 
all combinations which include informative augmentations
outperform CL4SRec, which demonstrates the necessity 
of using informative operators for short sequences.
With the additional `$\mathsf{M}$' augmentation, the performance
improves most. This is because `$\mathsf{M}$' operator shares the same information as the next-item prediction target, and it encourages the encoder to 
capture high-order item relationships. Additionally, the model with `$\{\mathsf{S},\mathsf{I},\mathsf{M}\}$'
operators outperforms `$\{\mathsf{S},\mathsf{I},\mathsf{M}, \mathsf{R},\mathsf{C}\}$'
indicating
that applying different augmentation operators to short and long sequences is meaningful as short sequences can be more sensitive to randomness.

\subsection{\textbf{Robustness Analysis (RQ3)}}

\begin{figure}[htb]
  \centering
  \includegraphics[width=\linewidth]{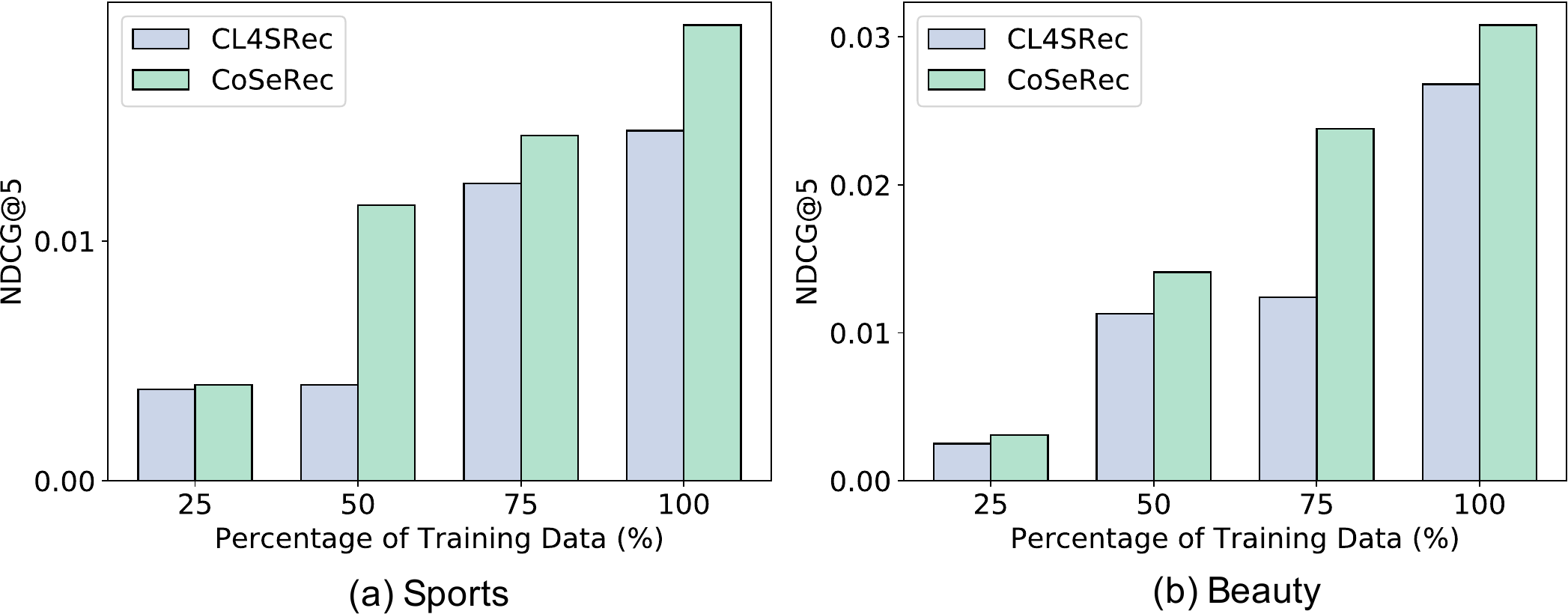}
  \caption{Performance comparison w.r.t. sparsity ratio.}
  \label{fig:robustness-of-proposed-method}
\end{figure}

Recommender systems usually suffer 
data-sparsity issues
where there are 
limited historical records.
To simulate this scenario, we
train a model with 
only partial training data 
(25\%, 50\%, 75\%, and 100\%) 
and keep the test data unchanged. 
We compare the proposed method
with the best baseline (CL4SRec)
on the Sports and Beauty datasets, which are presented 
in Figures~\ref{fig:robustness-of-proposed-method} (a) and (b), respectively.
We observe that 
performance
substantially 
degrades
when less training data is used, but \modelname consistently performs better than CL4SRec
and the performance degradation is slower than CL4SRec. For example, on Sports, \modelname 
achieves
similar performance with only 75\% of training data as of CL4SRec with full training data. Moreover, 
CL4SRec drops 76.19\% of its original performance while \modelname only drop 34\%, when both with 50\% training data.
These observations demonstrate that 
informative augmentations in \modelname
can the alleviate data-sparsity issue.
We also observe that the impact of data sparsity varies on different datasets. 
With 50\% training data,
\modelname's performance
drops 34\% on Sports while dropping 52.14\% on Beauty.

\begin{figure}[htb]
  \centering
  \includegraphics[width=\linewidth]{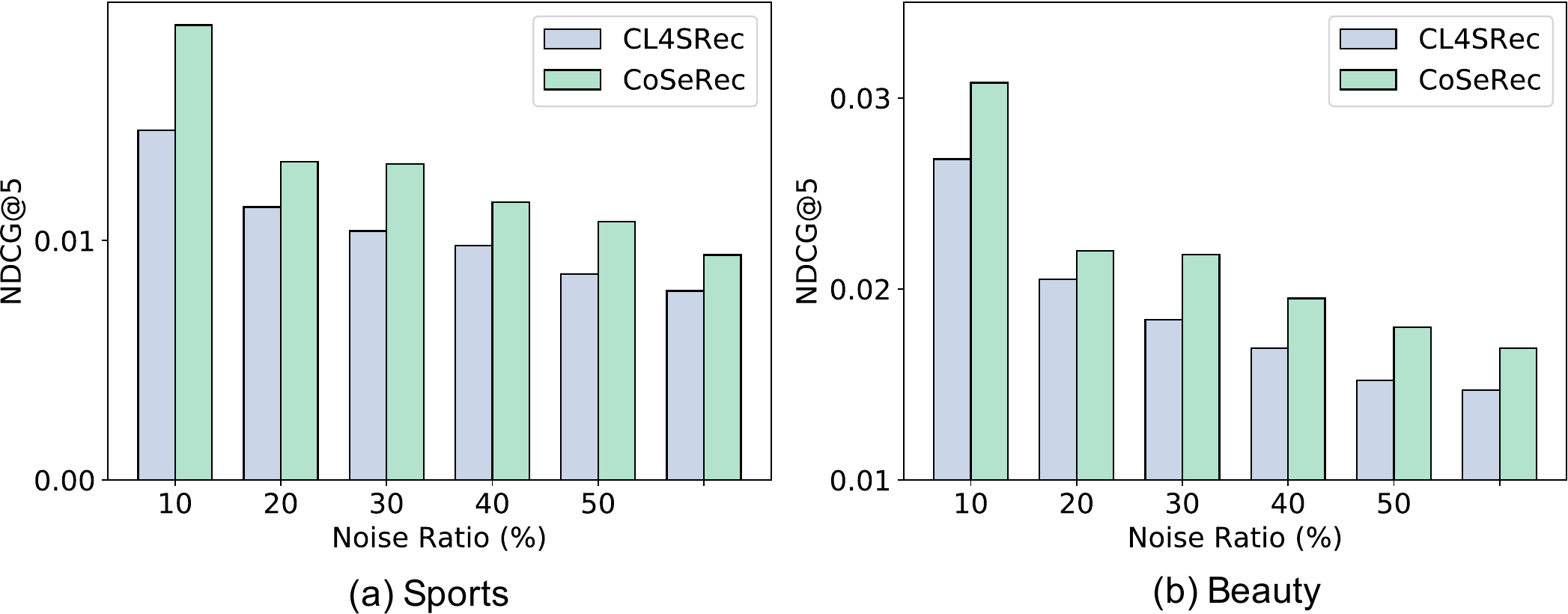}
  \caption{Model performance comparisons w.r.t. noise ratio. }
  \label{fig:robustness-noise-data}
\end{figure}

We also 
evaluate
the robustness of \modelname
to noisy interactions in 
the
inference phrase.
Specifically, we train a model 
with the original training data, 
and randomly add a certain 
proportion (10\%, 20\%, 30\%, 40\%, and 50\%) of 
negative user-item interactions
to every test sequence. Figures~\ref{fig:robustness-of-proposed-method} (c)-(d)
show the results on the Sports and Beauty datasets.
We see that
the 
performance
of both \modelname and CL4SRec 
decreases.
\modelname can consistently perform better than CL4SRec under any noise ratio.
This implies that with additional high quality augmentation operators (`Insert` and `Substitute`), \modelname
creates
higher confident positive views for the contrastive SSL objective to maximize agreements, which
thus endows the encoder with more robustness against the noisy interactions during the inference stage.

\subsection{\textbf{Study of \modelname (RQ4)}}

\begin{figure*}[htb]
  \centering
  \includegraphics[width=0.9\linewidth]{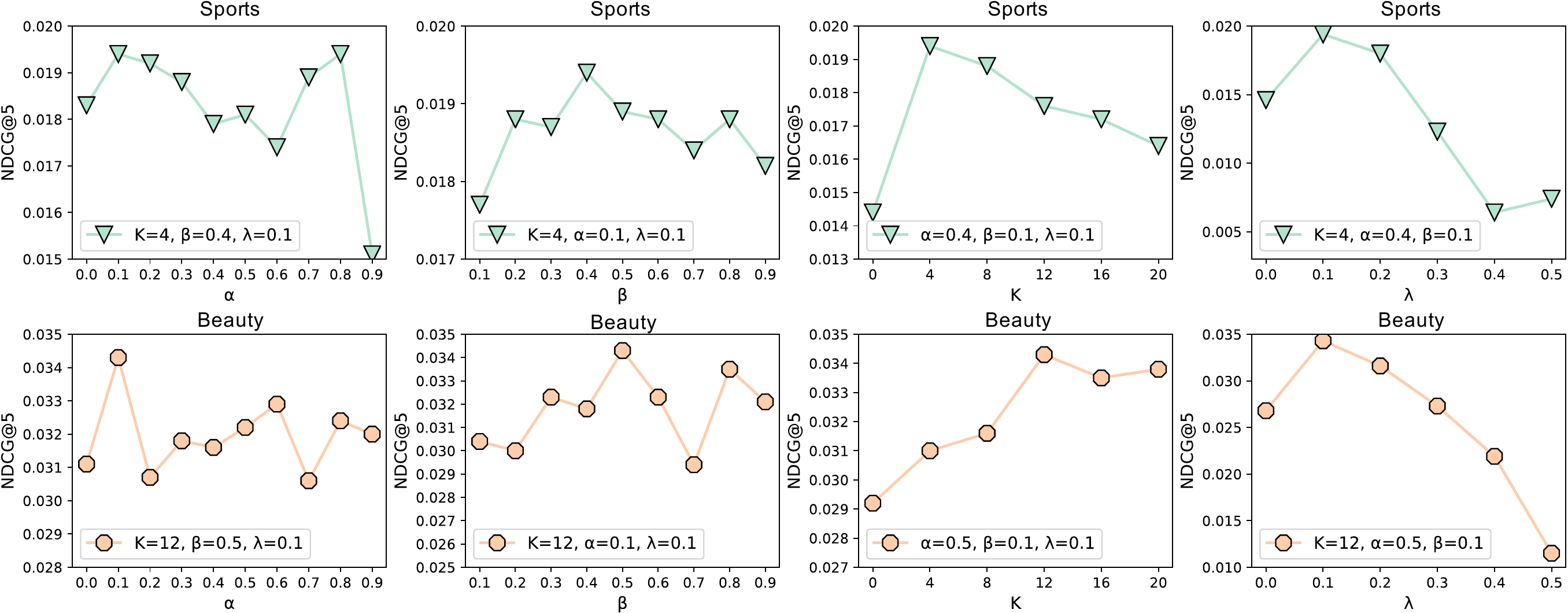}
  \caption{Performance comparison of \modelname w.r.t.~different $\alpha$, $\beta$, $K$, and $\lambda$ on Sports and Beauty datasets. 
  }
  \label{fig:all_hpo}
\end{figure*}

In this section,
we first explore the effects of hyperparamters $\alpha$,
$\beta$, $K$, and $\lambda$ in \modelname, 
which control the `Substitute' ratio, `Insert' ratio,
threshold of deciding whether a sequence to be short, and the intensity of contrastive SSL,
respectively.
Then we compare different item correlations.
Lastly, we investigate an alternative two-stages optimization strategy,
which first pre-trains from SSL and then fine-tunes on next-item prediction targets.

\subsubsection{\textbf{Hyperparameter Study.}}
\label{sec:hpo-sensitity}

We study the hyperparameters one by one meaning 
that all other hyperparameters are assigned
the optimal tuned values when studying a certain 
hyperparameter. Figure~\ref{fig:all_hpo} shows 
overall 
comparisons.

We 
observe that: 
(1) A general pattern for $\alpha$ and $\beta$
is that the performance reaches to a peak
as the ratios increasing
and then start to deteriorate.
The reason is that there is no informative augmentation when the ratio is $0$
and increasing the ratios too 
high
leads heavier corruptions, thus constructing
false positive samples to impair the SSL.
Specifically,
$\beta=0.4$ 
performs the best
while $\alpha=0.1$ 
performs the
best in 
the
Sports and Beauty datasets, respectively.
The larger $\beta$ 
indicates that
due to the long-tail distribution of short sequences, 
the dataset 
requires more `Insert' operator to alleviate the cold-start problem.
(2) 
The best values of $K$ are $4$ and $12$ 
on 
Sports and 
Beauty, respectively. 
We hypothesis that the difference between the optimal $K$ 
of two datasets is because
the length skewness is different.
(3) We can see that having the contrastive SSL signal ($\lambda > 0$)
is important as it significantly improves the performance 32.88\% and 12.82\% in $\mathrm{NDCG}@5$
on the Sports and Beauty datasets. We also observe that 
performance starts to deteriorate
when $\lambda > 0.1$. This indicates that the contrastive SSL signal
complements but should not dominate
the learning goal.

\subsubsection{\textbf{Effect of Item Correlations}}
We study the effect of different correlations 
introduced
in Section~\ref{sec:item-correlations} including 
\textit{memory-based}, \textit{model-based} and \textit{hybrid} correlations. Table~\ref{tab:item-correlations} shows
comparison results. 
We can see that using any type 
of
item correlations
helps improving 
performance
compared
with CL4SRec.
Among these correlations, \text{model-based} correlation
performs worse than \text{memory-based} and \text{hybrid}
correlations. This might 
be
because item representations
learned from 
the
model at early epochs are not informative.
\text{Hybrid} correlations 
performs
best with $E\geq 160$. This indicates that 
it is beneficial to incorporate 
both \text{model-based}
and \text{memory-based} correlations.

\begin{table}[htb]
\caption{Performance comparison w.r.t. different item correlations for
informative augmentations.}
\label{tab:item-correlations}
\begin{tabular}{l|cc|cc}
\toprule
\multirow{2}{*}{Correlation Type} & \multicolumn{2}{c|}{Sports} & \multicolumn{2}{c}{Beauty}\\
          & HR@5   & NDCG@5 & HR@5  & NDCG@5        \\ 
\hline
None (CL4SRec)  & 0.0231 & 0.0146 & 0.0401 & 0.0268 \\
\hline
Memory-based  & 0.0284 & 0.0194 & 0.0528 & 0.0357 \\
Model-based   & 0.0251 & 0.0164 & 0.0497 & 0.0338 \\
\hline
Hybrid ($E=80$) & 0.0270 & 0.0179 & 0.0503 & 0.0342\\
Hybrid ($E=160$) & 0.0287 & \textbf{0.0196} & \textbf{0.0537} & \textbf{0.0361}\\
Hybrid ($E=200$) & \textbf{0.0290} & 0.0194 & 0.0530 & 0.0357\\
\bottomrule
\end{tabular}
\end{table}

\subsubsection{\textbf{Effect of pre-training}}
\label{sec:pre-train-effect-and-insert-benefits}

We have demonstrated that jointly 
optimizing the recommendation objective 
with the contrastive learning objective
leads to an effective training of \modelname.
However, S$^3$-Rec claims that a \textit{two-stage} training, \textit{i.e.} pre-training with contrastive SSL and fine-tuning with next-item prediction, is also effective.
This two-stage training is also widely adopted in other areas~\cite{chen2020simple,devlin2018bert}.
Therefore, we conduct experiments to compare the performance of our multi-task training strategy and an alternative two-stage one.
Table~\ref{tab:join-fine-tune-with-data-aug} 
shows the result comparisons between 
multi-task training
and 
two-stage training on the Sports and Beauty datasets.
We observe that a two-stage training
performs worse
compared with the multi-task strategy.
The better performance of 
the multi-task strategy of \modelname reflects
that contrastive learning objectives 
and recommendation 
objectives can benefit from each other 
when jointly training, while two stage training
can lead to the pre-trained self-supervision information 
being forgotten in the
fine-tuning stage.

\begin{table}[htb]
\caption{Performance comparison under multi-task and two-stage training strategies on Sports and Beauty datasets.}
\label{tab:join-fine-tune-with-data-aug}
\begin{tabular}{l|cc|cc}
\toprule
\multirow{2}{*}{Strategy} & \multicolumn{2}{c|}{Sports} & \multicolumn{2}{c}{Beauty}\\
          & HR@5   & NDCG@5 & HR@5  & NDCG@5        \\ 
\hline
Multi-task & 0.0287 & 0.0196 & 0.0537 & 0.0361 \\
Two-stage  & 0.0242 & 0.0388 & 0.0435 & 0.0283 \\
\bottomrule
\end{tabular}
\end{table}

\section{Conclusion and Future Work}
In this work, we studied the contrastive SSL in sequential recommendation, which alleviate the data-sparsity and 
noisy interaction issues
We proposed a novel learning framework,
\modelname, which jointly optimizes the contrastive 
SSL objective and the next-item prediction objective.
We proposed two novel informative augmentation methods, \textit{i.e.} substitute and insert, which advances existing random augmentations by
leveraging
item correlations. Additionally, we endowed augmentations with sequence length awareness, which addresses the length skewness in sequential dataset. 
We conducted extensive experiments on three benchmark
datasets and justified the effectiveness and the robustness of \modelname. Moreover, we investigated the optimal augmentation for sequential recommendation, which verifies the efficacy of informative augmentations.

In the future, we plan to investigate 
more advanced model-based 
item correlation functions. For example,
searching 
for
correlated items under a
reinforcement 
learning framework so that more accurate 
correlation information can be leveraged. Additionally, we will explore finer-grained exploration of the relation between augmentations and the sequence length. Moreover, we 
plan to
study the effects of different contrastive learning functions in the self-supervised learning of sequences. 

\bibliographystyle{ACM-Reference-Format}
\bibliography{sample}
\appendix

\end{document}